\documentclass[12pt]{iopart}
\usepackage{graphicx} 
\usepackage{epstopdf}
\usepackage{color}
\usepackage[english]{babel}
\usepackage[parfill]{parskip}
\usepackage{siunitx}
\begin{document}

\title{Highly controlled optical transport of cold atoms into a hollow-core fiber}

\author{Maria Langbecker, Ronja Wirtz, Fabian Knoch, Mohammad Noaman, Thomas Speck, Patrick Windpassinger}

\address{Institut f\"{u}r Physik, Johannes Gutenberg-Universit\"{a}t Mainz, 55099 Mainz, Germany}
\ead{windpass@uni-mainz.de}
\vspace{10pt}
\begin{indented}
\item{May 15, 2018}
\end{indented}

\begin{abstract}
We report on an efficient and highly controlled cold atom hollow-core fiber interface, suitable for quantum simulation, information, and sensing. The main focus of this manuscript is a detailed study on transporting cold atoms into the fiber using an optical conveyor belt. We discuss how we can precisely control the spatial, thermal, and temporal distribution of the atoms by, e.g., varying the speed at which the atoms are transported or adjusting the depth of the transport potential according to the atomic position. We characterize the transport of atoms to the fiber tip for these different parameters. In particular, we show that by adapting the transport potential we can lower the temperature of the transported atoms by a factor of 6, while reducing the transport efficiency only by a factor 2. For atoms transported inside the fiber, we can obtain a transport efficiency into the fiber of more than 40\% and we study the influence of the different transport parameters on the time-dependent optical depth signal. When comparing our measurements to the results of a classical transport simulation, we find a good qualitative agreement. 
\end{abstract}

%
%
%
%

\section{Introduction}

Cold atoms are an ideal system for quantum simulation, computation, and sensing due to the high degree of control over their external as well as internal parameters. In the most common case, these atomic properties are manipulated by coupling the atoms to a light field. Thus, for a precise control, a well-defined atom-light interface is necessary. One possibility for an efficient atom-light interface are cold atoms inside a hollow-core fiber. Here, both light and atoms can overlap tightly throughout the length of the fiber and hence the interaction region can be several orders of magnitude larger than in free space. This also increases the optical depth $D_{\rm opt}$, which is a figure of merit for the effective light-matter interaction strength. Achievements with cold atoms in hollow-core fibers so far include ground-state electro-magnetically induced transparency \cite{Bajcsy2009, Duncker2014, Blatt2016}, used for example for an all-optical switch \cite{Bajcsy2009} and for light storage \cite{Blatt2016}, exciting and probing Rydberg atoms \cite{Langbecker2017}, precision spectroscopy \cite{Okaba2014}, and atom interferometry \cite{Xin2018}.

One key prerequisite for these applications is the controlled preparation of the atomic sample inside the hollow-core fiber. Different techniques have been demonstrated to load cold or ultracold atoms inside a hollow-core fiber, including the use of a single-beam red-detuned dipole trap as a guide \cite{Vorrath2010, Christensen2008}, a hollow beam blue-detuned dipole trap, free fall under gravity, a magnetic funnel \cite{Bajcsy2011}, a dark funnel in combination with a red-detuned dipole trap beam \cite{Blatt2014} or combinations of the above \cite{Xin2018, Hilton2018}. The most controllable way to transport atoms into a hollow-core fiber is to use a moving optical lattice, a so-called optical conveyor belt, first demonstrated by Okaba et al. \cite{Okaba2014}. A major advantage of this technique is that the atoms can be precisely transported and held at a specific position. This can for example enable a systematic survey of the inner part of the hollow-core fiber. Optical conveyor belts have previously been used to transport single particles \cite{Kuhr2001, Schrader2001} and Bose-Einstein-condensates \cite{Schmid2006}, for the study of coherence properties during the transport \cite{Kuhr2003} and to transport cold atoms outside an optical nanofiber \cite{Schneeweiss2012}. For long-distance transport in free space, Bessel beams \cite{Schmid2006} or movable lenses \cite{Middelmann2012} have been used to overcome limitations of the transport potential given by the Rayleigh range. 

While transporting the atoms into the hollow-core fiber, a typical problem is heating of the atoms \cite{Langbecker2017, Okaba2014, Xin2018,  Blatt2014}, for example due to their acceleration or due to the funnel shape of the transport potential, which increases towards the fibertip. However, low temperatures inside the fiber are advantageous, as they result e.g.\ in a longer free expansion time and in less heating-induced losses of atoms. To cool the atoms during the transport, so far Raman-sideband cooling \cite{Okaba2014} or continuous cooling in the magneto-optical trap until the atoms enter the fiber \cite{Blatt2014} have been employed.

In this manuscript, we introduce a method of controlling the atomic temperature by adapting the potential depth of the optical conveyor belt according to the position of the atoms. We transport the cold atoms over a distance of several millimeters towards and into our hollow-core fiber and characterize the influence of both frequency (acceleration) and amplitude (trap depth) ramps. By optimizing these transport settings, we show that we can realize a wide range of temperatures and corresponding particle numbers, which we compare to the results of a classical transport simulation. Further, we analyze the influence of the fiber on the transport process by studying measurements for atoms transported outside and inside the hollow-core fiber.

\section{Materials and Methods}
\subsection{Experimental Setup}
\begin{figure}
\center
\includegraphics{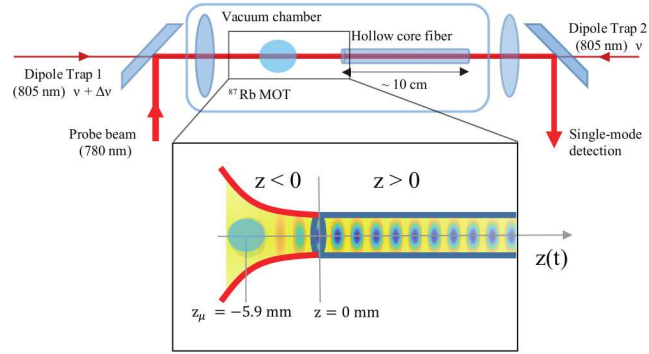}
\caption{\label{fig:setup} Schematic of the experimental setup. The main part shows a sketch of the vacuum system and the important beam paths, coupled through the hollow-core fiber. The inset shows a sketch of the optical lattice potential (lattice constant not to scale), which increases from the MOT position to the fiber tip and stays at constant depth inside the hollow-core fiber.}
\end{figure}

Our complete experimental setup has been described in detail in \cite{Langbecker2017}. \Fref{fig:setup} shows the details important for the transport procedure. We load a magneto-optical trap (MOT) of Rubidium 87 atoms at a distance of $|z_\mu |  = \SI{5.9}{\mm}$ from the tip of our hollow-core fiber. The fiber used in this work is a hollow-core photonic crystal Kagom\'e type fiber \cite{Couny2007} with a length of $\SI{10}{cm}$, a core diameter of $\SI{60}{\mu \m}$ and a mode field diameter of about $\SI{42}{\mu \m}$. At the MOT position, the cold atoms are loaded into a red-detuned optical lattice, which is created by two counter-propagating hollow-core-fiber-coupled trap beams at $\SI{805}{\nano \m}$. In addition to standard absorption imaging on a CCD camera outside the fiber, we have the possibility to detect the atomic absorption along the fiber axis using a resonant hollow-core-fiber-coupled probe beam.

\subsection{Highly controlled transport}
\label{sec:transport}

While loading the atoms into the optical lattice, the relative frequency detuning $\Delta \nu$ between the two dipole trap beams is set to zero. By increasing $\Delta \nu$, we create an optical conveyor belt to transport the atoms towards and into the fiber. The properties of this transport depend on the exact shape of the frequency ramp $\Delta \nu(t)$. In principle, arbitrary frequency ramps are possible in our setup as we can control the two dipole beam trap frequencies separately with two acousto-optical modulators which are driven by individual outputs of a programmable arbitrary waveform generator (FlexDDS by WieserLabs). We typically accelerate the atoms by ramping the frequency detuning of one of the beams linearly to a maximum value of a few hundred kilohertz. Before probing, we decelerate them by ramping the frequency detuning again down to zero.
Given the frequency detuning $\Delta \nu(t)$, we can determine the velocity of the moving lattice with wavelength $\lambda$, which corresponds to the transport velocity of the atoms, by
\begin{equation}
v(t)= \frac{\lambda}{2} \Delta \nu(t),
\label{eq:transport1}
\end{equation}
then integrate or differentiate this equation to find position and acceleration of the atoms. For the example of a linearly increasing frequency detuning, the final position of the atoms and the maximum acceleration during the ramp are given by 
\begin{equation}
z(\Delta t)=z_0+ \frac{\lambda \Delta \nu_{max} \Delta t}{4} \qquad {\rm and} \qquad a_{\rm max}=\frac{\lambda}{2} \frac{ \Delta \nu_{max}}{\Delta t}
\label{eq:transport2}
\end{equation}
respectively, where $z_0$ is the initial atomic position, $\Delta \nu_{max}$ the maximum applied frequency detuning and $\Delta t$ the duration of the ramp.
By choosing the appropriate frequency ramp, we can thus precisely control the atomic position, velocity and acceleration.   

\subsection{Adapting the potential depth}
\label{sec:amplitude_ramps}
Atoms in the optical lattice experience the trapping potential \cite{Grimm2000}
\begin{equation}
U(r,z,t) =  U_{\rm{dip}} \exp{\left(- \frac{2r^2}{\omega\left(z(t)\right)^2}\right)}\cos^2{\left( \pi \Delta \phi(t) - kz(t) \right)},
\label{eq:DT_potential}
\end{equation}
where $\omega\left(z(t)\right) = \omega_0 \left(1 + z(t)^2/z_R^2\right)^{1/2}$ is the waist of the Gaussian beam at position $z(t)$, with $z_R = \pi \omega_0^2/\lambda$ being the Rayleigh range and $\omega_0$ being the minimal beam waist at the fiber tip. The time-dependent phase shift $\Delta \phi(t)$ depends on the frequency ramp $\Delta \nu(t)$ as 
\begin{equation}
\Delta \phi(t) = \int \Delta \nu(t) dt,
\end{equation}
which in the easiest case of constant detuning $\Delta \nu$ reduces to $\Delta \phi(t) = \Delta \nu t$. The potential depth is given by
\begin{equation}
 U_{\rm{dip}}(z) = 2 \pi c^2 \left( \frac{2 \Gamma_{D2}}{\omega_{D2}^3 \Delta_{D2}} + \frac{\Gamma_{D1}}{\omega_{D1}^3 \Delta_{D1}}\right) I(z),
\label{eq:U_dip}
\end{equation}
where $\omega_{D1/D2}$ is the transition frequency from ground to excited state, $\Gamma_{D1/D2}$ is the decay rate from the excited state and $\Delta_{D1/D2}$ the detuning of the dipole trap beam from the transition frequency for the Rubidium D1 and D2 lines, respectively. The position-dependent maximum intensity of the dipole trap beams is given by
\begin{equation}
I(z) = \frac{2 P}{\pi \omega(z(t))^2},
\end{equation}
where $P$ is the power in each of the dipole trap beams. 

\begin{figure}
\center
\includegraphics{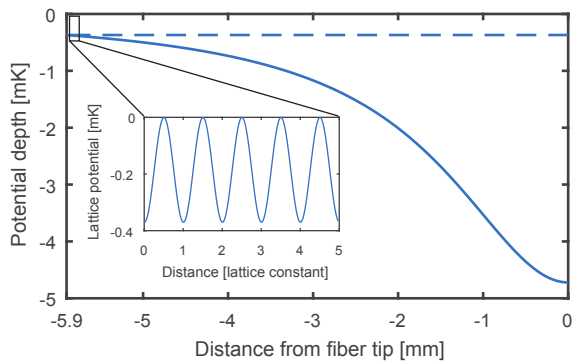}
\caption{\label{fig:potential} Potential depth (cf.\ \eref{eq:U_dip}) as function of distance from fiber tip for the case without laser intensity ramp down (solid line) and for the case of laser intensity ramp down with $n=1$ (cf.\ \eref{eq:U_AR}, dashed line). The inset shows the optical lattice created by the two counter-propagating dipole trap beams (cf.\ \eref{eq:DT_potential}).}
\end{figure}

\Fref{fig:potential} shows the behaviour of the trap depth $ U_{\rm{dip}}(z)$ between the MOT position and the fiber tip (main figure, solid line). The inset shows a zoom close to the MOT position to illustrate the lattice trapping potential $U(r=0,z,t=0)$. For $z<0$, the beam waist of the dipole trap beams, which have their focal point at the fiber tip, increases and thus intensity $I(z)$ and potential depth $U_{\rm{dip}}(z)$  decrease. We define the two limiting points as $U_{\rm{0}} =  U_{\rm{dip}}(z=z_\mu)$ for the potential depth at the MOT position and $U_{\rm{max}} =  U_{\rm{dip}}(z=0)$ for the maximum potential depth at the fiber tip. Inside the fiber for $z>0$, the trap depth will stay constant at $U_{\rm{max}}$, when neglecting the coupling efficiencies of the dipole trap beams into the fundamental mode of the hollow-core fiber. These are typically above $90 \%$ for our setup. For our typical experimental parameters, we find $U_{\rm{max}} \approx \SI{5}{mK}$ and $U_{\rm{0}} \approx \SI{400}{\mu K}$.

To load the atoms into the optical lattice, $U_{\rm{0}}$ needs to be much larger than typical temperatures in the MOT of tens of \SI{}{\mu K}, which imposes requirements on the trapping parameters. For example, for a far-detuned dipole trap at \SI{1064}{\nano \m}, achieving the same trap depth would require extremely high-power trapping beams of more than \SI{10}{W} per beam (cf.\ \eref{eq:U_dip}). To keep the laser powers moderate, we therefore choose to work with a near-detuned dipole trap at \SI{805}{\nano \m}. This, however, limits the lifetime of atoms in the dipole trap as will be discussed later.

Apart from this natural change in potential depth given by the Gaussian beam propagation, we can additionally modify the potential depth by changing the power $P$ of our dipole trap laser beams. We change the laser power in such a way that we adapt the potential depth to the position of the atoms, which inherently depends on time for a given frequency ramp (cf.\ \eref{eq:transport2}). The time-dependent potential including the amplitude ramp-down of the dipole trap beams has the form
\begin{equation}
 U_{\rm{dip}}(z,t) = n U_0 \left(1+A \left( \frac{z(t)}{z_R}\right)^2 \right) \left(1+ \left( \frac{z(t)}{z_R}\right)^2 \right)^{-1},
\label{eq:U_AR}
\end{equation}
with 
\begin{equation}
A = \frac{z_\mu^2 + (1-n) z_R^2}{nz_\mu^2}
\end{equation}
as a correction factor to maintain the overall shape of the potential. Here, $n$ controls the final potential depth as $U_{\rm{max}}= n U_0$. For example, if choosing $n=1$, the trap depth is being kept constant for all distances from the fiber tip and thus for the duration of the entire frequency ramp (see \fref{fig:potential}, main figure, dashed line).

In addition to this overall change in the dipole trap powers, we further can include individual corrections for each of the two dipole trap beam powers, e.g.\ to compensate for fiber coupling losses at high frequency detunings or for different initial power levels of the two dipole trap beams.

\subsection{Data acquisition and analysis}
\label{sec:data_analysis}
Outside the fiber, we characterize the atoms before and after the transport using standard absorption imaging on a CCD camera perpendicular to the fiber axis. By fitting a Gaussian density profile to the atomic cloud, we determine the number of atoms and the position of the atomic sample. We measure the radial temperature of the atomic cloud using the standard time-of-flight expansion method. 

For probing along the fiber axis and inside the hollow-core fiber, we use a circularly polarized probe beam resonant to the $^{87}\rm Rb$ transition $5 S_{1/2} (F=2) \rightarrow 5 P_{3/2} (F'=3)$ with an intensity much below the saturation intensity. This probe beam is coupled through the hollow-core fiber and then into a single-mode-fiber-coupled photomultiplier tube. With this second fiber-coupling, we select the part of the probe beam that has been guided in the fundamental mode of the hollow-core fiber. To ensure that we only measure atoms transported inside the hollow-core fiber, we apply a resonant push beam outside the fiber just before probing, which is perpendicular to the fiber axis.

From the frequency-dependent absorption profile of the probe beam, we determine the optical depth $D_{\rm opt}$ by a fit to the transmission $T = \exp\left(-D_{\rm opt}/\left[1+4\left(\Delta/\Gamma\right)^2\right] \right)$, where $\Delta$ is the detuning of the probe beam and $\Gamma$ the natural linewidth of the excited state. These spectroscopy measurements are performed in a time-resolved pulsed probing scheme, where the dipole trap is switched off for the duration of the probe pulses, as explained in detail in \cite{Langbecker2017}. To estimate the atomic temperature inside the fiber, we use a release-and-recapture fit, similar to e.g.\ \cite{Bajcsy2011}. We assume that during the dipole trap off time $t_{\rm off} =\SI{2}{\mu \s}$ the atomic cloud expands to $r_{\rm at}^2=r_0^2+v_{\rm at}^2 t_{\rm off} ^2$, where $r_0^2=\omega_0^2   (k_{\rm B} T)/U_{\rm dip}$ and $v_{\rm at}^2=2 (k_B T)/m$, with $T$ assumed to be constant throughout the probing process. We calculate the number of atoms recaptured by the dipole trap using the overlap integral of the probe beam intensity distribution $\exp{\left(⁡-2r^2/\omega_0^2\right)}$ with the atomic density distribution $n(r)=n_0 \exp{\left(-r^2/r_{\rm at}^2 \right)}$ inside the fiber with core radius $R_{\rm c}$. After $N$ of such release-recapture cycles, the optical depth is given by
\begin{equation}
\fl D_{\rm opt}(N) = \frac{N_{\rm at} \sigma_\pm S_{23}}{\pi \omega_0^2 } \left( \frac{2\omega_0^2}{2r_{\rm at}^2+\omega_0^2} \left( 1- \exp{⁡\left(-\frac{R_{\rm c}^2}{r_{\rm at}^2} \right)}   \exp{\left( ⁡-\frac{2R_{\rm c}^2}{\omega_0^2} \right)} \right) \right)^N,
\label{eq:release-recapture}
\end{equation}
where $\sigma_\pm$ is the resonant cross section and $S_{23}$ the hyperfine transition strength \cite{Steck2010}. We determine the initial atom number $N_{\rm at}$ for $N=1$ and $r_{\rm at}=r_0$.

\subsection{Theoretical simulations} 
\label{sec:theory}
To model the atomic transport, we assume all Rubidium atoms to be frictionless (very low temperatures) and independent, i.e.\ we neglect particle-particle interactions (low atom density). Thus, the single particle dynamics is solely governed by the dipole trap potential (cf.\ \eref{eq:DT_potential}), for which we solve Newton's equations of motion numerically by employing the implicit Adams integrator provided in the Python package SciPy~\cite{scipy}.

To prepare the initial state, we position 1500 particles at $x=y=\SI{0}{mm}$, while $z$-positions are drawn from a Gaussian distribution according to the experimental parameters. In a second step, every particle's $z$-position is shifted such that it coincides with its closest potential energy minimum. To initialize a starting temperature, the particle velocities are drawn from a Gaussian distribution with zero mean and width $\sigma_{v_x} =\sigma_{v_y} = \sigma_{v_z} = \left(k_B T_{\rm{init}}/m\right)^{1/2}$. Since all particles are positioned in potential energy minima, part of their average kinetic energy is converted into potential energy. Assuming a quadratic form around the potential minima and employing the equipartition theorem, the particle ensemble's equilibrated temperature is given by $\approx 1/2~T_{\rm{init}}$. After a short equilibration period of $\SI{20}{ms}$ we randomly select 1280 particles that did not escape the dipole trap, which are then used as the initial ensemble for the transport simulations. This number of particles is given by the available number of computational kernels.

To identify particles that have escaped the dipole trap, we monitor the total energy $E_i(t)$ of every particle individually. For $E_i(t)>0$ the particle escapes the trap and is no longer considered when computing ensemble averages. The loss of particles due to lifetime is taken into account at the final transport position. Finally, to estimate uncertainties of the quantities of interest, e.g., temperature and fraction of trapped particles, we split the full ensemble into 20 subensembles and compute the standard error.

\section{Results and Discussion}
In this section, we present a thorough characterization of our transport method and discuss the influence of the different transport parameters such as frequency and amplitude ramps on temperature and transport efficiency. First, we will concentrate on atoms transported to the fiber tip just outside the fiber (\Sref{sec:results_outside}) to characterize the transport itself, before investigating the influences of the hollow-core fiber by comparing the properties of atoms transported inside the fiber (\Sref{sec:results_inside}).

\subsection{Results outside the fiber}
\label{sec:results_outside}
\begin{figure*}
\includegraphics{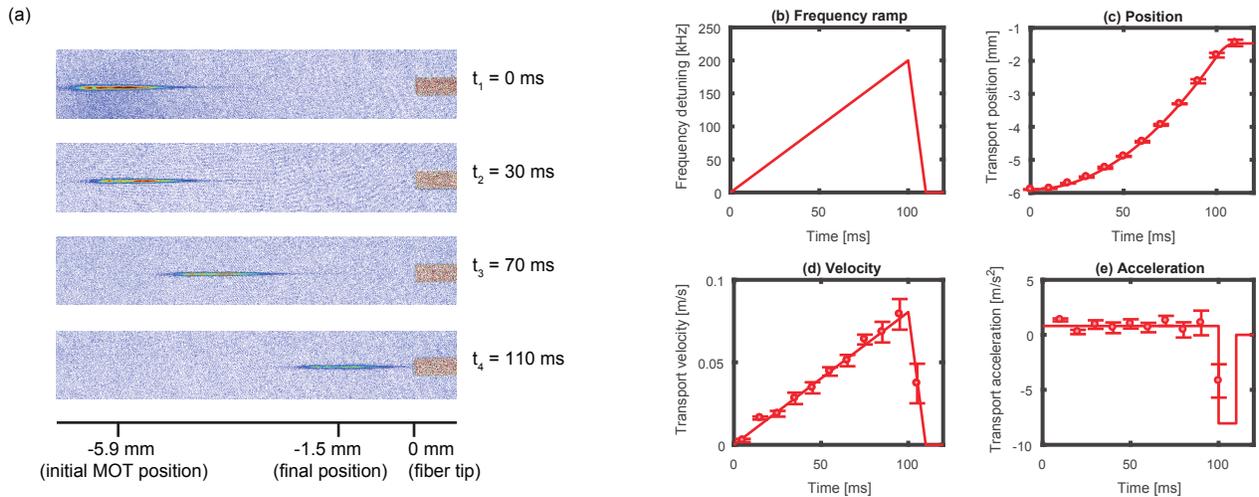}
\caption{\label{fig:transport} (a) Absorption images of an exemplary transport process. (c)-(e) Frequency detuning as well as atomic center of mass position, transport velocity and acceleration as a function of time. (Data points: Experimental data, solid lines: theoretical model (cf.\ \sref{sec:transport}). Error bars represent statistical errors.)}
\end{figure*}

To set the stage, we first demonstrate the transport process with a simple example. \Fref{fig:transport} (a) shows absorption images of an exemplary transport process for different time steps within the transport time. We determine the center-of-mass position $z(t)$ of the atomic cloud for each time step, which we differentiate to calculate velocity and acceleration during the transport. In this example, we use a linear frequency ramp $\Delta \nu(t)$ of $0$ to $\SI{200}{kHz}$ in $\SI{100}{ms}$ to accelerate the atoms and a linear frequency ramp of $200$ to $\SI{0}{kHz}$ in $\SI{10}{ms}$ for deceleration, as shown in \fref{fig:transport} (b). \Fref{fig:transport} (c)-(e) show both theoretical and experimental values for the position, velocity and acceleration of the atoms during this transport procedure. The agreement between experimental data and theoretical model confirms our high degree of control over the position of the atomic sample.

\subsubsection{Effect of frequency ramps on particle number}
\begin{figure*}
\includegraphics{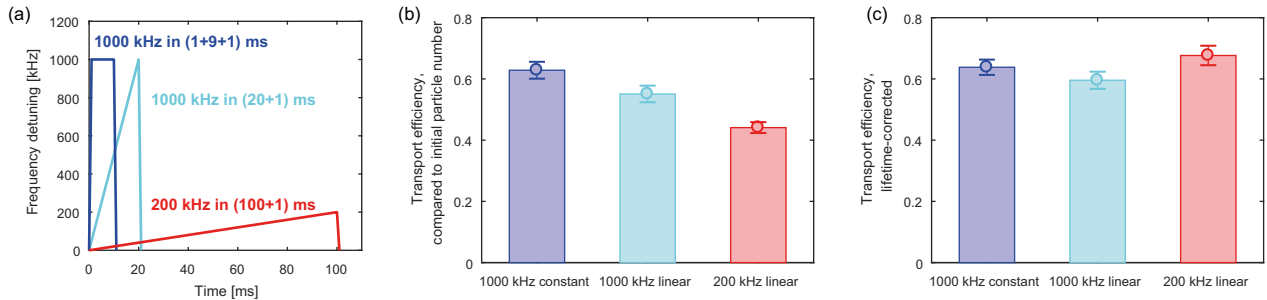}
\caption{\label{fig:particle_numbers}Transport efficiency for different frequency ramps. (a) Frequency detuning as function of time for different frequency ramps. (b) Transport efficiency compared to initial particle number. (c) Transport efficiency corrected for atom-loss due to finite lifetime in the dipole trap. (Error bars represent statistical errors.)}
\end{figure*}

For characterizing the transport process, we use different types of acceleration and deceleration ramps and discuss their influence on the transport efficiency without actively adapting the trapping potential. As a figure of merit, we use the number of transported atoms relative to the initial particle number loaded into the optical lattice at the MOT position. We detect the transported atoms directly after the end of the transport process without additional holding time or time of flight.
\Fref{fig:particle_numbers} (a) shows three different frequency ramps for accelerating the atoms. Two ramps have a linear increase in frequency detuning, which corresponds to a constant acceleration of the transported atoms as in the example shown in \fref{fig:transport}. The fastest ramp has a $\SI{1}{ms}$ frequency ramp-up and then a constant frequency detuning, which corresponds to a constant velocity of the transported atoms. We choose the timings for the different ramps according to the maximum frequency detuning in such a way that the atoms are always transported to the same final position $\SI{1.8}{mm}$ in front of the fiber tip. The velocity and acceleration for the different ramps (cf.\ \sref{sec:transport}) are given in \tref{tab:transport_parameters}. As we decelerate the atoms within $\SI{1}{ms}$ for all ramps, the magnitude of the deceleration is much larger than of the acceleration. However, even for the fastest ramp we are still far below the maximum acceleration $a_{\rm max}=U_0k/m=\SI{2.8e5}{m/s^2}$, for which the atoms are still transported in their potential wells \cite{Schrader2001}. Since we detect immediately after the transport, the transport efficiency is mainly influenced by the acceleration ramp.

\begin{table}
\caption{\label{tab:transport_parameters}Velocity and acceleration for different frequency ramps (cf.\ \sref{sec:transport}).}
\begin{tabular*}{\textwidth}{@{}l*{15}{@{\extracolsep{0pt plus
12pt}}l}}
\br
Frequency ramp&(Max.) Velocity&(Max.) Acceleration&Deceleration\\
\mr
\SI{1000}{kHz} constant&\SI[per-mode=symbol]{0.4}{\meter\per\second}&\SI[per-mode=symbol]{400}{\meter\per\square\second}&\SI[per-mode=symbol]{400}{\meter\per\square\second}\\
\SI{1000}{kHz} linear&\SI[per-mode=symbol]{0.4}{\meter\per\second}&\SI[per-mode=symbol]{20}{\meter\per\square\second}&\SI[per-mode=symbol]{400}{\meter\per\square\second}\\
\SI{200}{kHz} linear&\SI[per-mode=symbol]{0.08}{\meter\per\second}&\SI[per-mode=symbol]{0.8}{\meter\per\square\second}&\SI[per-mode=symbol]{80}{\meter\per\square\second}\\
\br
\end{tabular*}
\end{table}

We observe that the fastest ramp with $\SI{1000}{kHz}$ constant detuning has the highest transport efficiency (\Fref{fig:particle_numbers} (b)). The efficiency decreases as the ramp time increases, that is for linear ramps and lower maximum detunings. This effect is due to the finite lifetime of atoms held in the dipole trap, which is for example limited by scattering of the dipole trap beams, resonant light from the dipole trap tapered amplifiers \cite{Savard1997} and the background vacuum pressure. To correct our transport efficiencies for the finite lifetime in the trap, we compare the number of transported atoms with the atom number at the initial position after the same holding time (see \fref{fig:particle_numbers} (c)). Here, we observe that all studied frequency ramps show a similar lifetime-corrected transport efficiency, with the slowest ramp showing the highest value. That means that within our experimentally possible range of frequency detunings, the lifetime has the largest effect on the transport efficiency. Thus, for the highest absolute final atom number, constant detuning ramps at high detunings are the best choice for our experimental parameters due to their fast ramping speed. They are for example better suited for transporting atoms further into the fiber. However, for higher detunings we observe a tail of atoms trailing behind the transported atoms, which corresponds to atoms lost from the transport process. They may be the reason for the lower lifetime-corrected transport efficiency for the two high detuning ramps. Thus, for experiments not limited by lifetime, slower ramps might be preferable. We further note that all our transport ramps are fundamentally limited by a transport efficiency of about 70\%, even when taking the lifetime correction into account. Limiting factors could be amplitude and phase noise on our dipole trap beams, as these have been shown to have an impact on long-distance transport \cite{Schmid2006}.

\subsubsection{Effect of amplitude ramps on temperature}
\begin{figure*}
\includegraphics{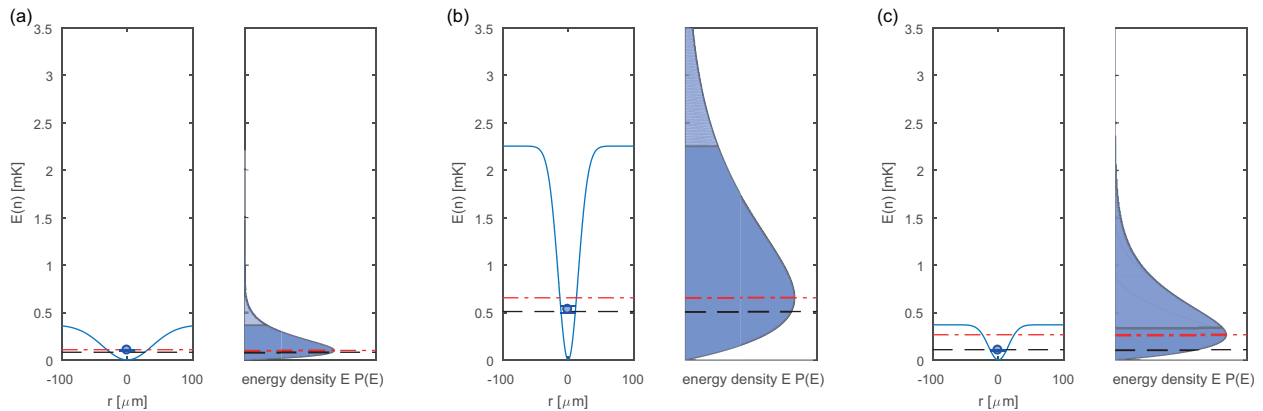}
\caption{\label{fig:temperatures} Experimental and theoretical temperatures of atoms in the dipole trap (a) at MOT position, (b) after transport to the fiber tip without amplitude ramp-down, (c) after transport to the fiber tip with amplitude ramp-down ($n=1$). On the left-hand side of each figure the trapping potential in radial direction is plotted with the experimentally measured temperature (data point) as well as the theoretical mean temperature (red dashed-dotted line) and the temperature corresponding to the truncated Boltzmann distribution (black dashed line). On the right-hand side of each figure, the energy density $EP(E)$ corresponding to the theoretical mean temperature is plotted. The dark-coloured area indicates where the Boltzmann distribution is truncated by the trapping potential.}
\end{figure*}

Next, we characterize the effect of our transport procedure and the trapping potential on the temperature of the atomic cloud. After loading the atoms into the dipole trap, we measure a temperature of $T \approx \SI{100}{\mu K}$. This temperature corresponds to a specific energy level in the trapping potential, which we approximate as a two-dimensional harmonic oscillator in the radial direction (see \fref{fig:temperatures} (a), left-hand side). The energy density $EP(E)$ (see \fref{fig:temperatures} (a), right-hand side) is then given by the Boltzmann distribution $P(E)=\exp{\left(-E/k_B T\right)}/Z$, where $Z$ is the partition function \cite{Reif1965}.

After being transported to the fiber tip, the atoms will experience an increased dipole trap potential $U_1$ (see \fref{fig:temperatures} (b), left-hand side). Due to this, the trapping frequency of the harmonic oscillator will have increased and the same energy level will have a higher energy and corresponding higher temperature $T_1$. Thus, if we assume entropy to stay constant during the transport so that the atoms still occupy the same energy level, the mean atomic temperature will also have increased (red dashed-dotted line in \fref{fig:temperatures} (b)). However, for this given mean temperature, not all the atoms will remain trapped. By using a so-called truncated Boltzmann distribution \cite{Tuchendler2008}, we calculate the mean temperature of the remaining atoms (black dashed line in \fref{fig:temperatures} (b)) as
\begin{equation}
\langle T_1 \rangle _{2D,trunc}= T_1 \frac{1 - \left(1+\eta+\frac{1}{2} \eta^2 \right) \e^{-\eta}}{1 - \left(1+\eta \right) \e^{-\eta}},
\end{equation}
where $\eta=  U_1/\left(k_B T_1\right)$. This value of about $\SI{510}{\mu K}$ is in good agreement with the value of $\SI{530 \pm 38}{\mu K}$  measured experimentally for the \SI{1000}{kHz} linear ramp (blue data point in \fref{fig:temperatures} (b)).

When we keep the potential depth constant by applying an amplitude ramp-down, the atoms which are transported to the fiber tip will nevertheless experience a change in trapping potential profile as due to the smaller beam waist the shape of the potential will be compressed (see \fref{fig:temperatures} (c), left-hand side). That leads to a large part of the Boltzmann distribution to be “squeezed” out of the trap (see \fref{fig:temperatures} (c), right-hand side), thus losing about 60\% of the transported atoms. The mean temperature calculated from the truncated Boltzmann distribution of about $\SI{110}{\mu K}$ is much lower than the temperature obtained for transport without amplitude ramp-down and close to the initial temperature. Again, the theoretical calculation is in good agreement with the experimentally measured value of $\SI{100 \pm 7}{\mu K}$.

We can thus control the atomic temperature by adapting the potential depth using amplitude ramps of the dipole trap laser beams. This comes at the cost of losing atoms when aiming for low final temperatures.

\subsubsection{Effect of combined frequency and amplitude ramps}
\begin{figure*}
\includegraphics{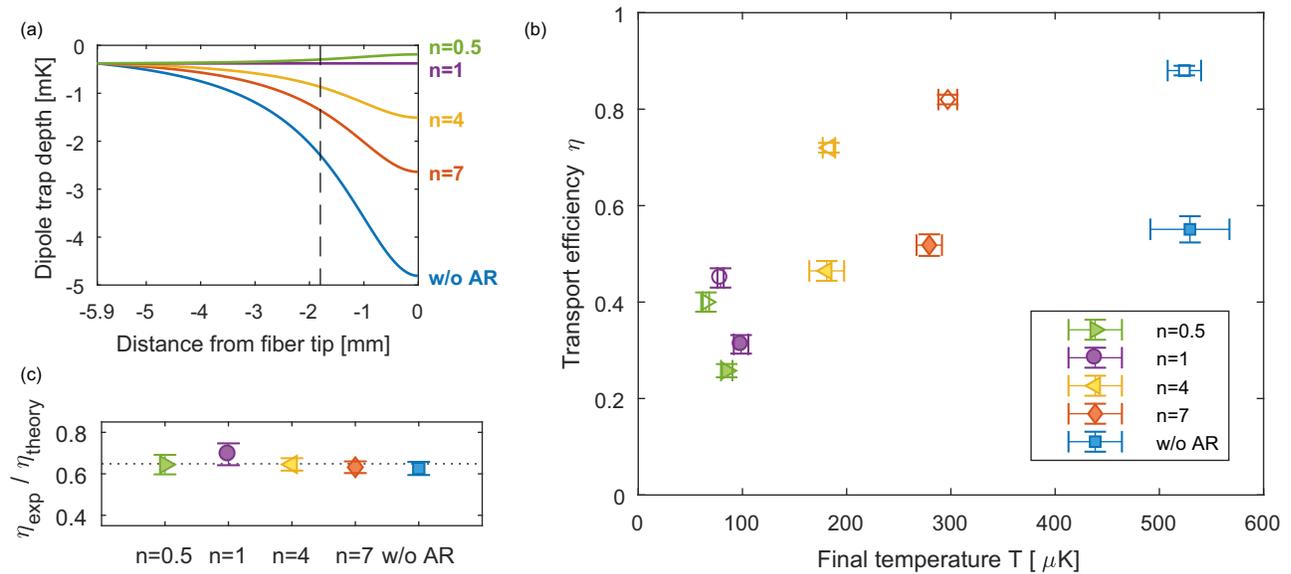}
\caption{\label{fig:comparison_outside} (a) Potential depth as function of distance for different amplitude ramp-downs. The black dashed line marks the final atomic position for the applied frequency ramp. (b) Experimental (filled markers) and theoretical (empty markers) results for transport efficiency and temperatures for atoms transported to the fiber tip for the \SI{1000}{kHz} linear frequency ramp and for different amplitude ramp-downs. (c) Ratio of experimental and theoretical values for the transport efficiency. The black dotted line marks the mean value for all different amplitude ramps. (Error bars represent statistical errors and errors from the fitting procedure.)}
\end{figure*}

Finally, we study the combined effect of different frequency and amplitude ramps on the transported atoms. We exemplarily show the results for the \SI{1000}{kHz} linear frequency ramp for amplitude ramps with different final dipole trap depths, as shown in \fref{fig:comparison_outside} (a). For this, we measure both the transport efficiency compared to the initial atom number and the final temperature of the atomic cloud (\fref{fig:comparison_outside} (b), solid data points). As discussed before, we achieve high transport efficiencies up to 55\% for high final dipole trap depths, but at the same time the temperatures are much higher than the initial temperature. When lowering the final trap depth, both efficiency and final temperature decrease. For a final trap depth as low as $4U_0$, we still observe a transport efficiency of about 45\%. However, the final temperature is reduced by more than a factor of 3. For very low final trap depths, the transport efficiency reduces to about 25\%. Still, these ramps are a good choice if one is interested in obtaining a very low-temperature final atomic sample, with temperatures even below the initial temperature and more than 6 times lower than the temperatures without amplitude ramp. 

We have confirmed that also the other two frequency ramps shown in \fref{fig:particle_numbers} (a) exhibit qualitatively the same behavior as discussed above and that the qualitative overall trend for each ramp is robust with regards to day-to-day performance fluctuations of the experiment.

The results from the theoretical simulation (\sref{sec:theory}) are plotted in \fref{fig:comparison_outside} (b) (empty data points) together with the experimental results. We note a very good agreement regarding the qualitative behavior of the different ramps. When comparing the experimental and theoretical values for the temperatures, we see a very good agreement for high final dipole trap depths. However, for small dipole trap depths, we measure a higher temperature than theoretically expected. This can be due to fluctuations in laser powers as discussed above, which most strongly influence lower trap depths. When comparing the experimental and theoretical values for the transport efficiency, we find a constant correction factor of around 0.65 for the \SI{1000}{kHz} linear ramp (see \fref{fig:comparison_outside} (c)). In the same way, we determine a correction factor of around 0.75 for the \SI{200}{kHz} linear ramp. This confirms our previous observation of the limited maximum lifetime-corrected transport efficiencies (cf.\ \fref{fig:particle_numbers} (c)), where the efficiency for the \SI{200}{kHz} linear ramp was also higher than for the \SI{1000}{kHz} linear ramp. A constant correction factor further implies that the additional experimental loss mechanism does not depend on the number of transported atoms, which would be consistent with loss due to laser noise as discussed above. Overall, we find that our model gives a good understanding of the transport mechanism and predicts the results for the individual ramps well, except for correction factors for experimental imperfections.

\subsection{Results inside the fiber}
\label{sec:results_inside}
When transporting the atoms inside the hollow-core fiber, our figure of merit to characterize the transport is the time-dependent optical depth $D_{\rm opt}$, which is proportional to the number of transported atoms, but additionally depends e.g.\ on the overlap between atomic cloud and probe beam (cf.\ \sref{sec:data_analysis}). Due to a micro lensing effect of the atomic cloud \cite{Roof2015, Gilbert2018}, we find that calculating the optical depth for high atomic densities is not straight forward. The details of this investigation will be discussed elsewhere \cite{Noaman2018}. For the measurements to be discussed in the following, we have therefore transported the atoms far enough (\SI{6}{mm}) inside the fiber that we expect these lensing effects to play a minor role due to reduced atomic densities.

\begin{figure*}
\includegraphics{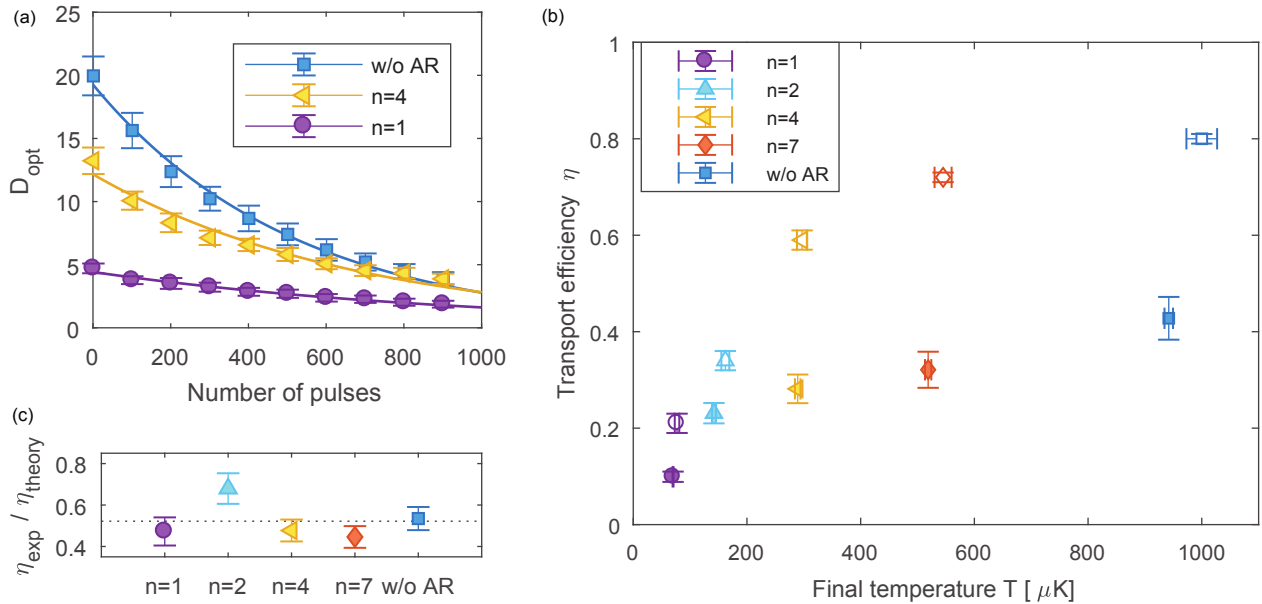}
\caption{\label{fig:comparison_inside} (a) Decay of the optical depth with number of probe pulses for different amplitude ramp-downs (data points) and corresponding release-and-recapture fit (\Eref{eq:release-recapture}, solid line). (b) Experimental (filled markers) and theoretical (empty markers) results for transport efficiency and temperatures for atoms transported 6 mm inside the fiber for the \SI{1000}{kHz} constant frequency ramp for different amplitude ramp-downs. (c) Comparison of experimental and theoretical values for the transport efficiency. The black dotted line marks the mean value for all different amplitude ramps. (Error bars represent errors from the fitting procedure.)}
\end{figure*}

\Fref{fig:comparison_inside} (a) shows the behavior of the optical depth during the pulsed probing process (cf.\ \sref{sec:data_analysis}) for three different final dipole trap depths for the \SI{1000}{kHz} constant frequency ramp. We observe that the initial optical depth is highest for the transport without ramping down the potential and decreases for lower final trap depths. This corresponds to our expectations from studying the transport efficiencies outside the fiber. We further observe that with increasing number of probe pulses the optical depth decreases. In principle, this behavior occurs for all amplitude ramps, but is most prominent for high final trap depths, for which the atomic cloud has a higher temperature. We can determine this temperature using a release-and-recapture fit (cf.\ \sref{sec:data_analysis}), which is also plotted in \fref{fig:comparison_inside} (a).

\Fref{fig:comparison_inside} (b) shows both the transport efficiency compared to the initial atom number and the final temperature of the atomic cloud for amplitude ramps with different final dipole trap depths for the \SI{1000}{kHz} constant frequency ramp (solid data points). Here, the results inside the fiber are qualitatively very similar to the results we have obtained outside the fiber. Without amplitude ramp-down, we obtain both higher transport efficiencies and higher temperatures, while when ramping down the amplitude to lower final trap depths, both transport efficiency and temperatures decrease. Also the overall shape of this decrease agrees well with the measurements outside the fiber. Our transport efficiencies range from more than 40\% for an atomic ensemble at about $\SI{940}{\mu K}$ to about 10\% for a very cold atomic ensemble at about $\SI{70}{\mu K}$. Thus, by adapting the trap depth, we can lower our final temperatures by more than a factor of 10, while only decreasing the transport efficiency by a factor of 4.

As for the measurements outside the fiber, we note a very good agreement between experimental results and the results from the theoretical simulation (\sref{sec:theory}, shown in \fref{fig:comparison_inside} (b) as empty data points) regarding the qualitative behavior of the different ramps. When comparing the experimental and theoretical values for the transport efficiency, we again find a constant correction factor (see \fref{fig:comparison_inside} (c)). With a value of about 0.5, it is lower than the correction factor measured outside the fiber. One reason for this is that the experimental limitations for the transport as observed outside the fiber have a stronger impact for longer transport distances. Another effect the theory does not consider is influences from the fiber itself other than keeping the trapping potential constant. Thus, a lower experimental transport efficiency points to an experimental limit when loading the atoms into the fiber. We expect the main reason for this to be imperfect coupling of the dipole trap beams into the fundamental mode of the hollow-core fiber as discussed in \sref{sec:amplitude_ramps}, as we have observed that this overlap strongly influences the transfer efficiency of atoms from outside to inside the fiber. This effect leads to a lower effective trap depth, which can also accommodate for the experimental values for the temperatures lying systematically below the theoretical values.

With our transport without amplitude ramp-down, we calculate a loading efficiency from the MOT into the hollow-core fiber of about 2.7\%, which is very comparable to values reported in other groups (2.5\% in \cite{Blatt2014} and 3\% in \cite{Hilton2018}). We also would like to point out that this number represents a lower limit for our loading efficiency, as we measure atoms transported already \SI{6}{mm} inside the fiber.

\section{Conclusion}
\label{sec:conclusion}
In conclusion, we have presented a detailed study on the transport of cold atoms both inside and outside a hollow-core fiber using an optical conveyor belt. We have found that by applying optimized frequency and amplitude ramps we can control a wide range of transport efficiencies and final atomic temperatures. In particular, we can prepare final atomic ensembles at the same  temperature as before the transport. Depending on the individual experimental requirements, we can find a compromise between the final number of atoms and their temperature. While for atoms with a higher mean temperature, we can achieve higher absolute transport efficiencies, colder atomic ensembles show an extended observation time before colliding with the fiber walls, which is advantageous for in-fiber experiments. We further find that our experimental results agree qualitatively well with the theorectical simulation and we have a good understanding of the correction factors between experiment and theory. Our high degree of control over the atom-fiber interface will for example allow for experiments to systematically survey the inner part of the hollow-core fiber. We also expect our results to be of importance for other conveyor belt or atoms-in-fiber experiments for reducing heating during transport. 

\ack
We are very grateful to Florian Stuhlmann who set up and programmed the FlexDDS during his bachelor's thesis. We thank Parvez Islam and Wei Li for helpful discussions and we further thank the GPPMM group of Fetah Benabid for design and production of the hollow-core fibers for our experiment. We gratefully acknowledge financial support by the DFG SPP 1929 GiRyd, the DFG via the collaborative research center TRR 146 (Project No. A7) and FP7-PEOPLE-2012-ITN-317485 (QTea).

\section*{References}
\bibliography{bibliography}{}

\providecommand{\newblock}{}
\begin{thebibliography}{10}
\expandafter\ifx\csname url\endcsname\relax
  \def\url#1{{\tt #1}}\fi
\expandafter\ifx\csname urlprefix\endcsname\relax\def\urlprefix{URL }\fi
\providecommand{\eprint}[2][]{\url{#2}}

\bibitem{Bajcsy2009}
Bajcsy M, Hofferberth S, Balic V, Peyronel T, Hafezi M, Zibrov A~S, Vuletic V
  and Lukin M~D 2009 {\em Phys. Rev. Lett.\/} {\bf 102} 203902

\bibitem{Duncker2014}
Duncker H 2014 {\em {Ultrastable Laser Technologies and Atom-Light Interactions
  in Hollow Fibers}\/} Ph.D. thesis Universit\"at Hamburg

\bibitem{Blatt2016}
Blatt F, Simeonov L~S, Halfmann T and Peters T 2016 {\em Phys. Rev. A\/} {\bf
  94} 043833

\bibitem{Langbecker2017}
Langbecker M, Noaman M, Kj\ae{}rgaard N, Benabid F and Windpassinger P 2017
  {\em Phys. Rev. A\/} {\bf 96} 041402(R)

\bibitem{Okaba2014}
Okaba S, Takano T, Benabid F, Bradley T, Vincetti L, Maizelis Z, {Yampol'skii}
  V, Nori F and Katori H 2014 {\em Nat. Commun.\/} {\bf 5} 4096

\bibitem{Xin2018}
Xin M, Leong W~S, Chen Z and Lan S~Y 2018 {\em Sci. Adv.\/} {\bf 4} e1701723

\bibitem{Vorrath2010}
Vorrath S, M{\"o}ller S, Windpassinger P, Bongs K and Sengstock K 2010 {\em
  \NJP\/} {\bf 12} 123015

\bibitem{Christensen2008}
Christensen C~A, Will S, Saba M, Jo G~B, Shin Y~I, Ketterle W and Pritchard D
  2008 {\em Phys. Rev. A\/} {\bf 78} 033429

\bibitem{Bajcsy2011}
Bajcsy M, Hofferberth S, Peyronel T, Balic V, Liang Q, Zibrov A~S, Vuletic V
  and Lukin M~D 2011 {\em Phys. Rev. A\/} {\bf 83} 063830

\bibitem{Blatt2014}
Blatt F, Halfmann T and Peters T 2014 {\em Opt. Lett.\/} {\bf 39} 446--449

\bibitem{Hilton2018}
Hilton A~P, Perrella C, Benabid F, Sparkes B~M, Luiten A~N and Light P~S 2018
  {\em arXiv:1802.05396\/}

\bibitem{Kuhr2001}
Kuhr S, Alt W, Schrader D, M{\"u}ller M, Gomer V and Meschede D 2001 {\em
  Science\/} {\bf 293} 278--280

\bibitem{Schrader2001}
Schrader D, Kuhr S, Alt W, M\"uller M, Gomer V and Meschede D 2001 {\em Appl.
  Phys. B\/} {\bf 73} 819--824

\bibitem{Schmid2006}
Schmid S, Thalhammer G, Winkler K, Lang F and {Hecker Denschlag} J 2006 {\em
  \NJP\/} {\bf 8} 159

\bibitem{Kuhr2003}
Kuhr S, Alt W, Schrader D, Dotsenko I, Miroshnychenko Y, Rosenfeld W,
  Khudaverdyan M, Gomer V, Rauschenbeutel A and Meschede D 2003 {\em Phys. Rev.
  Lett.\/} {\bf 91} 213002

\bibitem{Schneeweiss2012}
Schneeweiss P, Dawkins S~T, Mitsch R, Reitz D, Vetsch E and Rauschenbeutel A
  2012 {\em Appl. Phys. B\/} {\bf 110} 279--283

\bibitem{Middelmann2012}
Middelmann T, Falke S, Lisdat C and Sterr U 2012 {\em \NJP\/} {\bf 14} 073020

\bibitem{Couny2007}
Couny F, Benabid F, Roberts P~J, Light P~S and Raymer M~G 2007 {\em Science\/}
  {\bf 318} 1118--1121

\bibitem{Grimm2000}
Grimm R, Weidem\"uller M and Ovchinnikov Y~B 2000 Optical dipole traps for
  neutral atoms ({\em Advances In Atomic, Molecular, and Optical Physics\/}
  vol~42) ed Bederson B and Walther H (Academic Press) pp 95 -- 170

\bibitem{Steck2010}
Steck D~A 2010 Rubidium 87 d line data available online at
  http://steck.us/alkalidata (revision 2.1.4, 23 December 2010)

\bibitem{scipy}
Jones E, Oliphant T, Peterson P {\em et~al.\/} 2001-- {SciPy}: Open source
  scientific tools for {Python} [Online; accessed 07.05.2018]
  \urlprefix\url{http://www.scipy.org/}

\bibitem{Savard1997}
Savard T~A, O'Hara K~M and Thomas J~E 1997 {\em Phys. Rev. A\/} {\bf 56}
  R1095--R1098

\bibitem{Reif1965}
Reif F 1965 {\em {Fundamentals of Statistical and Thermal Physics}\/} (New
  York: McGraw-Hill)

\bibitem{Tuchendler2008}
Tuchendler C, Lance A~M, Browaeys A, Sortais Y~R~P and Grangier P 2008 {\em
  Phys. Rev. A\/} {\bf 78} 033425

\bibitem{Roof2015}
Roof S, Kemp K, Havey M, Sokolov I~M and Kupriyanov D~V 2015 {\em Opt. Lett.\/}
  {\bf 40} 1137--1140

\bibitem{Gilbert2018}
Gilbert J~R, Roberts C~P and Roberts J~L 2018 {\em J. Opt. Soc. Am. B\/} {\bf
  35} 718--723

\bibitem{Noaman2018}
Noaman M, Langbecker M and Windpassinger P 2018 Micro lensing induced
  lineshapes in a single mode cold atom hollow-core fiber interface (in
  preparation)

\end{thebibliography}

\end{document}